\documentclass[apj]{emulateapj}
\usepackage{hyperref}
\usepackage{graphicx}
\usepackage{amssymb}
\shortauthors{Osten \& Crosley}

\begin{document}
\title{Quantifying the ngVLA's Contribution to Exo-space Weather: Results of a Community Studies Report
\textit{Next Generation VLA Memo \# 31}}
\shorttitle{Winds, Particles, and Fields from Stars }

\author{Rachel A. Osten\altaffilmark{1}}
\affil{Space Telescope Science Institute}
\email{osten@stsci.edu}
\author{Michael K. Crosley}
\affil{Physics \& Astronomy Dept., Johns Hopkins University, Baltimore, MD }

\altaffiltext{1}{Also at Center for Astrophysical Sciences, Johns Hopkins University, Baltimore, MD 21218}

\vspace*{1cm}

\begin{abstract}
A forward-looking facility such as the next generation Very Large Array (ngVLA) requires forward-looking science. 
The ngVLA will enable stellar wind detections or robust constraints on upper limits sufficient to bridge the gap
between current radio upper limits on cool stellar mass loss and current indirect estimates of stellar mass loss from astrospheric
measurements. This will aid in understanding the interplay between the rotational spindown of stars  
and wind mass loss rates, important for an understanding not just of stellar astrophysics but the impact on 
planet formation and planetary dynamos.
Radio observations are currently the only way to explore accelerated particles
in cool stellar environments. 
The upgraded sensitivity of the ngVLA will enable the detection and characterization of essentially all classes of
radio-active cool stars in the galaxy. This will open up new fields of study and contribute to the understanding of
the stellar contribution to exo-space weather, especially with the emphasis on sensitivity above a frequency of about 10 GHz,
where the transition to optically thin gyrosynchrotron emission occurs.
These topics connect and extend the key science topics in the cradle of life, in an area which there is both high academic 
as well as popular interest, and which is expected to continue into the next several decades.
\end{abstract}

\section{Introduction}
With the announcement by the Kepler mission of more than two thousand confirmed transiting exoplanets, 
population statistics suggesting that a transiting exoplanet in the habitable zone of its star likely resides within
5 pc (Dressing \& Charbonneau 2013), 
and recent confirmation via radial velocity techniques
of a planet in the habitable zone of our nearest star (Anglada-Escud\'{e} et al. 2016),
astronomy is firmly in the age of exoplanets. 
This is an area which grabs the public's attention as well as
unites professional astronomers, planetary scientists, and astrobiologists, in the search for life. 
Ground- (MEarth project) and space-based (TESS, PLATO) projects have as their primary goal the discovery
of even more, closer transiting exoplanets,
and large ground- and space-based projects like the James Webb Space Telescope and giant segmented
mirror telescopes (E-ELT, TMT, GSMT) will characterize their atmospheres.
Characterizing the atmospheres and environments of potentially habitable exoplanets will be a major focus of the astronomical
community in the next ten to twenty years.  
During that time, we can expect that the phase space for exoplanet demographics will 
be fleshed out, with major inroads into understanding the interrelationship between planet properties and stellar properties.
There will be many potentially habitable worlds, and one major question will be how to move beyond that potentiality into
a more informed assessment of the likelihood of seeing an \textit{inhabited} world.

Stellar activity is known to be able to create false positive signatures in biomarkers (Tian et al. 2014).
Planetary habitability is a multi-faceted issue and depends not just on the planetary parameters, but is strongly
influenced by the environment set by the host star.  
The star can influence the planetary environment through 
radiation and particles.  Stellar flaring and associated coronal mass ejections, a steady stellar wind, and star-planet
magnetospheric interactions are all important factors in the space weather environments of these exoplanets.
Radio emission is the only wavelength regime capable of placing direct constraints on the particle environment that
stars produce, and adds significant information on the steady mass loss of nearby stars.

\section{Mass Loss on the Lower Half of the Main Sequence}
Stellar winds affect the migration and/or evaporation of exoplanets (Lovelace et al. 2008), and are 
important not only for understanding stellar rotational evolution, but it also their influence
on planetary dynamos (Heyner et al. 2012).
At present only indirect measures of cool stellar mass loss inform these topics.  
Mass loss 
in the cool half of the HR diagram, along/near the main sequence, has been notoriously difficult to detect directly, 
due in part to the 
much lower values of mass loss here compared to other stellar environments 
($\dot M$ of $2\times$ 10$^{-14}$ solar masses per year for the Sun). 
Detection of Lyman $\alpha$
astrospheric absorption (Wood et al. 2004) does not detect the wind directly, but rather the bow shock created when the
wind interacts with the local interstellar medium. This can only be done with high-resolution ultraviolet spectrographs in space; currently
the Space Telescope Imaging Spectrograph on the Hubble Space Telescope, installed in 1997, is the only instrument
capable of making such measurements. Non-detections of this feature do not provide upper limits to the stellar mass loss. 

Cool stellar mass loss is characterized by an ionized stellar wind, whose radio flux can have a $\nu^{0.6}$ or
$\nu^{-0.1}$ dependence if in the optically thick or thin regime, respectively.  
A direct measurement of stellar 
mass loss through its radio signature would be a significant leap forward not only for understanding the plasma physics of the 
stars themselves, but also for understanding what kind of environment those stars create. 
Previous attempts at a direct detection of cool stellar mass loss via radio emission have led to upper limits typically
 three to four orders of magnitude higher than the Sun's present-day mass loss, while indirect methods find  evidence
for mass loss rates comparable to or slightly higher than the Sun's present day mass loss rate (up to $\sim$80 times solar
$\dot M$.)

Initial work assessing the feasibility of the ngVLA's ability to detect directly the stellar winds of nearby cool, main
sequence stars was done as part of the Bower et al. (2015) report of ngVLA Working Group 4.
This brief work used analytic formulae in Drake et al. (1987), based on the expected optically thick and thin
radio emission from an ionized stellar wind presented in Wright \& Barlow (1975), and Panagia \& Felli (1975); optically thick emission
from an ionized stellar wind is expressed as \\
\begin{equation}
S_{\nu} = 1.6\times10^{11} \left( \frac{\dot{M}}{v_{w}} \right)^{1.33} \frac{\nu^{0.6}_{5} T_{4}^{0.1}}{d_{kpc}^{2}} 
\end{equation}
where $S_{\nu}$ is the predicted radio flux in mJy per beam, $\dot{M}$ is the mass loss rate in solar masses per year,
$v_{w}$ is the velocity of the stellar wind, $\nu_{5}$ is the frequency normalized to 5 GHz, $T_{4}$ indicates the
temperature of the solar wind in units of 10$^{4}$K, and $d_{kpc}$ is the distance in kpc.  As winds from late-type stars are presumed to originate
in the open field lines of the outer corona, a coronal wind temperature of 10$^{6}$ or $T_{4}$=100 is used.
We expand on these calculations to different frequencies in the expected ngVLA range, and incorporate demographics
of the nearest stars using catalogues of nearby stars.  We use the recent continuum sensitivities quoted for the ngVLA at a variety of frequencies
as posted at \url{http://science.nrao.edu/futures/ngvla/concepts}, extended for a 12 hour integration. 
The stellar escape velocity is $\sim$600 km s$^{-1}$ for main sequence solar-like stars and cooler,
and the minimum constrainable mass loss rate for this velocity is quoted in this memo. However,
we note that recent results
from studies of exoplanets (Vidotto et al. 2015 and others) have found evidence for hot Jupiters embedded in slower stellar winds
than would be expected based on solar scalings, with expectations of a denser stellar wind than experienced by solar system planets. With this in mind,
we explore a range of assumed wind speeds, from 100 to 1200 km s$^{-1}$.

\subsection{Increasing the Look-back Time for Studying the Young Sun}

Table~1 lists nearby commonly used solar analogs within 15 pc from the Sun, and their ages.  
Results from high spectral resolution UV data of nearby stars where astrospheric absorption is seen
suggest that there is a power-law relationship between inferred mass loss and surface X-ray flux, with a
limit in surface X-ray flux of about 10$^{6}$ erg cm$^{-2}$ s$^{-1}$, above which no or significantly reduced mass loss is inferred.
This limits the ability to constrain the mass-loss history of the Sun, using analogues. In particular, the lack of detection of this feature
does not provide any quantitative constraints on the mass-loss of that star. 
Based on this work, the most active stars appear not to have strong winds;  there is evidence via  astrospheric detection towards the nearby young
solar analog $\pi^{1}$UMa (Wood et al. 2014), with age estimates between 300 and 500 million years, of a weak wind from the young Sun.

There is tension between the results returned from the astrospheric detection method and the mass loss expected from stellar rotational evolution models.
Stars are expected to lose angular momentum in a wind as their rotation decreases with time; the rate of rotation period decrease can be determined from distributions
of rotation periods in clusters of known ages,
 while MHD wind models are required to interpret the implied mass loss rates which should accompany this.  Johnstone et al. (2015)
performed such calculations, suggesting that the stellar mass-loss rate can be parameterized as \\
\begin{equation}
\dot{M_{\star}} \propto R_{\star} \Omega_{\star}^{1.33} M_{\star}^{-3.36} \; \; .
\end{equation}
The results suggest that the Sun at young ages should have had a mass loss rate roughly an order of magnitude higher than it does today.
Fichtinger et al. (2017) presented recent results obtained from the JVLA and ALMA for a sample of nearby stars. Results are shown in Figure~1. They expanded upon the earlier spherically symmetric
wind models, and showed that a conical opening angle for the wind can reduce the mass loss rate implied from a given radio flux density by about a factor of two 
compared to the spherically symmetric case. Their upper limits for implied mass loss are still two to three orders of magnitude above the levels of mass loss implied from
the rotational evolution models, which themselves are about an order of magnitude above the mass loss rate implied by the detection of astrospheric absorption
from the nearby solar analogue $\pi^{1}$ UMa. Results from the ngVLA will be able to make a significant leap in understanding the mass loss history of the Sun
through its two order of magnitude improvement over what can be achieved at present with the JVLA. In addition, non-detections provide quantitative constraints
on the mass loss rate, a marked difference from the astrospheric absorption method.

\begin{table}
\caption{Commonly used Solar Analogs and Potential ngVLA Mass-Loss Constraints}
\begin{tabular}{llll}
Name & Age & Distance & $\dot{M}$ constraint$^{1,2}$ \\
           & (MY) & (pc) & ($\dot{M}_{\odot}$) \\
\hline
EK Dra & 100 & 34 & 425\\
$\chi^{1}$ Ori & 300 & 8.7 & 55 \\
$\pi^{1}$ UMa & 500 & 14.3 & 116 \\
$\kappa^{1}$Cet & 650 & 9.2 & 60 \\
\hline
\multicolumn{4}{l}{$^{1}$ Using ngVLA sensitivity at 17 GHz,} \\
\multicolumn{4}{l}{extrapolated for a 12 hour integration} \\
\multicolumn{4}{l}{$^{2}$ For $\dot{M}_{\odot}$=2$\times10^{-14}$ M$_{\odot}$ yr$^{-1}$ and wind} \\
\multicolumn{4}{l}{speed equal to escape speed} \\
\end{tabular}
\end{table}

\begin{figure}
\hspace*{-1.5cm}
\includegraphics[scale=0.4]{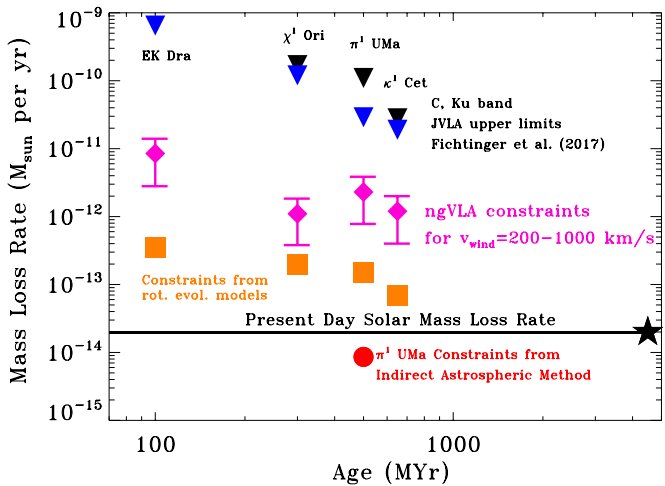}
\caption{Summary of current mass loss constraints for nearby solar analogs, along with prospects achievable with the ngVLA. Black and blue
triangles indicate upper limits from C and Ka bands obtained with the JVLA (Fichtinger et al. 2017) for spherically symmetric winds; orange squares show mass loss constraints from the rotational 
evolution models of Johnstone et al. (2015); and red circle shows detection from the indirect method of inferring mass loss via astrospheric absorption (Wood et al. 2014).
Magenta diamonds and error bars display the grasp of the ngVLA, for wind velocities spanning 200-1000 km s$^{-1}$. The present-day solar mass loss rate is 2$\times$10$^{-14}$
$\dot{M}$ yr$^{-1}$.}

\end{figure}

\subsection{M Dwarfs in the Solar Neighborhood}
After initial skepticism about the possibility of M dwarfs hosting exoplanets, attention is fully fixed on these cool dim stars. 
M dwarfs are the most common type of star in our galaxy, and recent results have shown
a high occurrence rate of planets around M dwarfs.  This makes
planets around M dwarfs one of most common modes of planet formation
in our galaxy.
The magnetic activity of M dwarfs can have significant influence on the chemistry and dynamics of these planets' atmospheres due to the planets' close
proximity to the star, and these effects
may be observable with JWST (Venot et al. 2016).
Recent results (Garaffo et al. 2016, 2017) have shown that the stellar magnetosphere influences the inner edge of the traditional habitable zone. In the case of the
iconic TRAPPIST-1 system of seven planets, magnetospheric models suggest that all but two of these planets would have orbits crossing the 
Alfv\'{e}n surface and thus would experience severe space weather. 
Thus consideration of the steady stellar wind as well as time-varying flares and coronal mass ejections from M dwarfs are vital components to understanding
the complex intertwining that may or may not turn a potentially habitable planet into an inhabited planet.
Current studies suggest that there are about 270 M dwarfs within 10 pc (Figure ~2). Several of these are already known to host exoplanets; by the time that the ngVLA
is operational, at the end of the 2020s, all of these will have have been surveyed for both close-in exoplanets (via the transit method), and those in more distant 
orbits (by coronography). Radio observations to constrain the mass loss from a steady stellar wind will be crucial measurements to add to the mix to understand
the impact the star may have on exoplanet companions.

\begin{figure}
\hspace*{-1.5cm}
\includegraphics[scale=0.4]{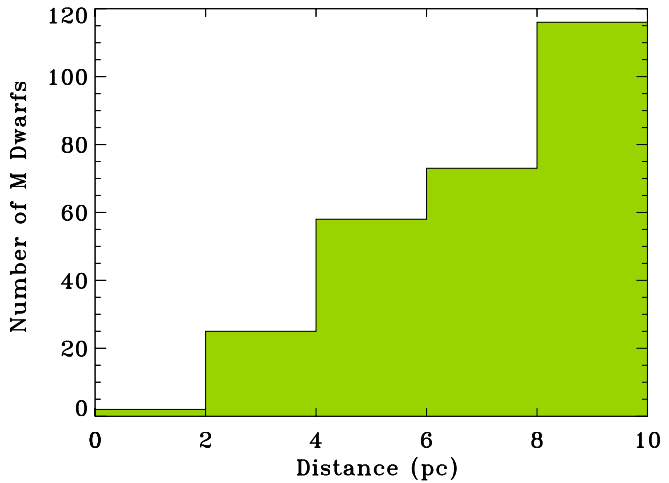}
\caption{Number of M dwarfs as a function of distance, from the REsearch COnsortium of Nearby Stars (RECONS). There are more than about 270 M dwarfs within 10 pc;
several of these are known to host exoplanets and by the time of the late 2020s all of these will have been surveyed for close-in exoplanets. An important constraint
for further characterization of these exoplanets will be the extent to which the star makes its proximate environment helpful or harmful for habitability.
}
\end{figure}

Very little is known observationally about the mass-loss of nearby M dwarfs. The nearby, magnetically active M dwarf EV~Lac is one exception, having had an 
 astrospheric detections from Lyman $\alpha$ absorption.
The inference is for a mass loss rate of only 0.7 times the solar mass loss rate, after correcting the results of Wood et al. (2004) for the much smaller surface area
of EV~Lac. 
Wargelin \& Drake (2001) presented a different method for detecting stellar winds, which makes use of an X-ray charge exchange halo
which should exist between the highly charged ions in the stellar wind and the surrounding ISM. This method is promising; they were able to provide an
upper limit to the mass loss rate of Proxima at 3$\times$10$^{-13}$ M$_{\odot}$ yr$^{-1}$, with approximately a factor of three uncertainty in their model.
This requires the use of a space-based X-ray emission with sufficient sensitivity and spatial resolution to resolve this emission from the stellar coronal emission, expected
to be thousands of times stronger (the spatial resolution is required to discern this signature in the wings of the coronal PSF).
This may be a complementary approach to the detection of radio emission of an ionized stellar wind, possible with the Lynx mission concept under
study by NASA with 50 times Chandra's sensitivity and Chandra-like angular resolution.  Because of the angular resolution requirement, it is most competitive
for stars within about 5 pc.

The method of detecting radio emission from nearby M dwarfs will be an important contributor to understanding how the wind environment of
nearby M dwarfs contributes to the habitability of orbiting exoplanets. Figure~3 shows the sensitivity of the ngVLA to detecting radio emission from an ionized stellar wind for M dwarfs within 10 pc,
using similar methods as described in the above section for the mass loss history of the Sun.  The
green triangle depicts the constraint on mass loss rate for the nearest M dwarf, Proxima, described above, along with the estimated uncertainty in the model. 
The red circle shows the mass loss rate in solar masses per year for the nearby M dwarf EV~Lac, as obtained from the detection of astrospheric
absorption.

\begin{figure}
\hspace*{-1.5cm}
\includegraphics[scale=0.4]{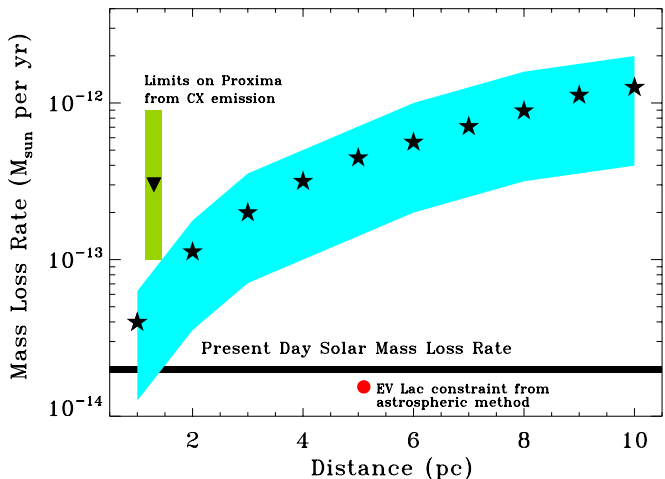}
\vspace*{-1.5cm}
\caption{Regions of stellar mass loss rate which can be constrained by a 12 hour ngVLA integration for studies of radio emission from an ionized stellar wind, for M dwarfs at a range of distances
from 1-10 pc. Stars indicate expected amount of optically thick emission at 28 GHz from a spherically symmetric wind with wind velocity equal to the
escape velocity, and cyan shaded region gives area encompassed by 
a range of assumed wind velocities, from 200 - 1000 km$^{-1}$.
Current observational limits on M dwarf wind mass loss are indicated at the distance of the objects: the green triangle indicates the upper limit for Proxima based on limits
from charge exchange emission (Wargelin \& Drake 2001), along with the uncertainty in their model. The stellar wind mass loss rate implied by the detection of
Lyman $\alpha$ astrospheric absorption towards the M dwarf EV~Lac at 5.1 pc is also indicated.
}
\end{figure}

\subsection{Interpretation of Radio Emission}
While measurements of radio emission from point sources are easy to undertake, their interpretation is complicated. 
The calculations in the above sub-sections have been performed under the assumption of optically thick emission, in which case the 
emission increases as $\nu^{0.6}$. The commonly observed gyrosynchrotron emission from radio active stars
typically has a peak frequency near 10 GHz, with flat or slightly negative spectrum at higher frequencies. Indeed, the upper limits
returned by Fichtinger et al. (2017) were obtained after careful removal of observed variable emission attributed to gyrosynchrotron emission.
The peak frequency is a function of activity, so 
observations at higher frequencies are necessary to separate any wind emission from these other sources.  At even higher frequencies such as those probed by ALMA,
stellar chromospheric emission is detected.  Thus the region between 10 and 100 GHz is ideal for searching for wind emission from nearby stars.

The stellar wind must be optically thin to nonthermal radio emission originating closer to the stellar surface. 
The time-averaged mass-loss from coronal mass ejections, considered a component of a stellar wind, 
can not be higher than a steady spherically symmetric stellar wind. Both Drake et al. (2013) and Osten \& Wolk (2015) have argued for a
high flare-associated transient mass loss from active stars; for the young solar analog EK~Dra flare-associated CME-induced mass-loss
rate of 4$times$10$^{-11}$ M$_{\odot}$ yr$^{-1}$.  The mass loss rate of M dwarfs is expected to vary widely as a function of magnetic activity,
with flare-associated mass loss rates of 10$^{-12}$ and 10$^{-11}$ M$_{\odot}$ yr$^{-1}$ for the active M dwarfs AD~Leo and EV~Lac, respectively, both at distances of about
5 pc. Constraints from ngVLA measurements would be below both of these numbers, and would provide 
a robust constraint on the breakdown of flare-CME connections, as implied in Crosley et al. (2017).
Instrument requirements are a large enough bandwidth to
disentangle different contributions to the spectral energy distribution and mitigate variability expected to dominate in the lower frequency bands,
and circular polarization for additional constraints on ruling out nonthermal emission.

\section{Particles and Fields: Contribution to Exo-Space Weather}
Habitability studies are interested in the flux of accelerated particles directed outward from the stellar surface; the most energetic ones
are the most impactful to an exoplanetary atmosphere. Studying the population of accelerated particles
near the stellar surface can provide details of the particle spectrum. Comparison to well-studied solar events enables
systematics with respect to scalings or extrapolations used for stellar space weather calculations to be investigated.
Segura et al. (2010) investigated the impact on the ozone layer of a superflare from an M dwarf. The energetic particles can 
deplete the ozone layer of a planet in the habitable zone of an Earthlike planet around an M dwarf experiencing a superflare. 

Radio observations provide the clearest signature of accelerated particles and shocks in stars arising from transient magnetic reconnection, and provide more realistic constraints on these factors than scaling by solar values (Osten \& Wolk 2017). Optically thin gyrosynchrotron emission constrains the index of the accelerated particles, and 
also enables a constraint on magnetic field strengths in the radio-emitting source (Smith et al. 2005, Osten et al. 2016). The peak frequency is about 10 GHz, so observations
at higher frequencies are necessary to disentangle optical depth effects from the interpretation.

\subsection{Radio-Active Cool Stars}
Variable stellar radio emission in the cool half of the HR diagram is more typical than steady flux levels, at least at microwave frequencies. This is a consequence of
the dominance of emission from accelerated particles, and an indicator of the nonsteady levels of particle acceleration, varying as
a function of position on the stellar disk and time. While previous studies estimating the contribution of stellar magnetic reconnection flaring
from different types of active stars have used average or ``typical" flux densities, the existence of extreme levels of variability
(up to 1-2 orders of magnitude) make this approach susceptible to biases. The timescales for stellar
flares can vary from milliseconds, in the case of some highly circularly polarized radio bursts (Osten \& Bastian 2006, 2008),
to minutes to many hours or days for the most dramatic examples (Richards et al. 2003). 


We examined objects characteristic of their class for several types of radio-active cool stars illustrating a large amount of variability.
This included Algol, RS CVn, and BY Dra active binaries, and nearby magnetically active mid-M and late-M dwarf stars.
We used a radio luminosity distribution
which gives the range of radio emission expected to be observed for each type of object, under the assumption that the flux variation seen in 
one object is typical of the class. Figure~4
shows this time-domain information recast as a cumulative probability distribution for each object.
The time-domain information for Algol, HR 1099, $\sigma^{2}$CrB, and UX~Ari is taken from multi-year
monitoring with the Green Bank Interferometer  (Richards et al. 2003). The M dwarf distribution
comes from extensive 8 GHz light curves of the nearby M dwarf EV~Lac (Osten et al. 2005, Osten et al. in prep.),
while the ultracool dwarf LP349-25 was used for the distribution of radio luminosities for ultracool dwarfs
(Ngoc et al. 2007, Osten et al. 2009).  For young stellar objects, the radio luminosity distribution was 
obtained from the 556 sources detected in the deep centimeter wavelength catalog of the Orion Nebula Cluster
(Forbrich et al. 2016). For this sample, objects with negative spectral index were removed
as potential contaminating extragalactic sources, and the distribution of radio luminosities for the 
remaining 478 sources was computed.


We used demographic information and space densities (listed in Table~2) to come up with a likelihood of observing a given radio luminosity level for
types of stars at a given distance. This is more probabilistic and represents reality better than taking the largest radio
luminosity of a given type, assuming that that is characteristic level of radio emission, and projecting that to find the
farthest distance to which the system is sensible, to arrive at source numbers. An rms for given frequency and integration time determines the distance to
which an observation is sensitive, for a source of a particular radio luminosity, \\
\begin{equation}
L_{R} = F_{\rm 3\sigma} 4 \pi d_{\rm sens}^{2}
\end{equation}
with $L_{R}$ the radio luminosity, $F_{\rm 3\sigma}$ the flux density density sensitivity, and $d_{\rm sens}$ the resulting
sensitivity distance.  Then, the number of stars is 
\begin{equation}
\# Stars = \left( \frac{f_{\rm sky}}{4\pi} \right) \frac{4 \pi n_{\rm obj}} {3} \left( \frac{L_{R}}{4\pi F_{3\sigma}} \right)^{3/2} P(L_{R})
\end{equation}
where $f_{\rm sky}$ is the fraction of the sky visible from the ngVLA site, $\approx$ 10.3 steradian (Condon et al. 1998),
$n_{\rm obj}$ is the space density of the type of star, L$_{R}$ the radio luminosity, F$_{3\sigma}$ the
sensitivity threshold, and $P(L_{R})$ the probability of the stellar object having that luminosity.
We realize that the assumption of spherical  geometry will break down at the scale of the Milky Way thick disk, or ~1 kpc, but this treatment gives a simplified approach to the maximum distance detectable.
Figure~5 plots the number of stars detectable with the ngVLA as a function of radio luminosity, using the radio luminosity 
distributions shown in Figure~4.
Only luminosity ranges available from the data shown in Figure~4 are used, and inflections in the radio 
luminosity distribution show up as departures from the $\propto L_{R}^{3/2}$ trend of number of stars as a function
of radio luminosity in Figure~5. The active binary systems have larger intrinsic radio luminosities
than the M dwarf classes.

Figure~6 shows the number of stars detectable as a function of distance. Here the distance at which the
trend for each type of star flattens indicates the distance at which the smallest radio luminosity
present in Figure~4 is no longer detected. All stars are detectable at small distances, and
the number scales as distance$^{3}$. Eventually only the brighter stars are able to be detected
at farther distances, so the rate of additional star counts decreases until no more stars would be
detected at the largest distances. The calculation for M dwarfs used two different space density values as
listed in Table~2.

\begin{table}[h]
\caption{Categories of radio-active cool stars and references to their space densities.}
\begin{tabular}{lll}
\hline
Category & Space Density & Reference \\
	& stars pc$^{-3}$ & \\
\hline
active binaries & 3.7$\times$10$^{-5}$ & Favata et al. (1995) \\
early-mid M dwarfs & 0.08 & Reid et al. (2007) \\
					& 0.05 & Reid et al. (2008) \\
ultracool dwarfs & 0.013 & Reid et al. (2008) \\
\hline
\end{tabular}
\end{table}

\begin{figure}
\hspace*{-1cm}
\includegraphics[scale=0.45,trim=0 0 50 200,clip]{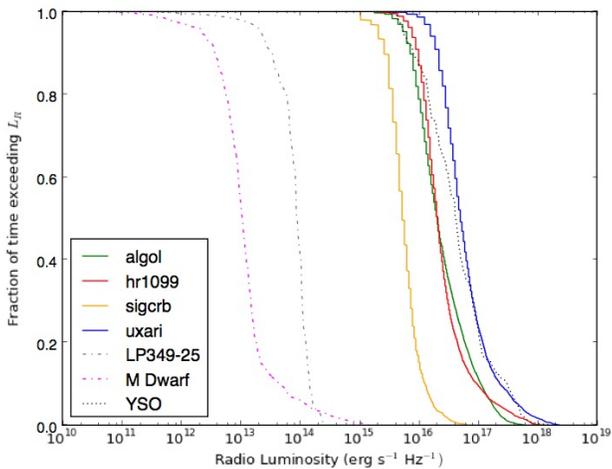}
\vspace*{-3cm}
\caption{ Cumulative distributions of 8 GHz cm radio luminosities for different categories of variable cool stellar objects: red, blue, yellow,
and green histograms are active binary systems (Algol and RS~CVn-type systems); the purple
dashed histogram is the result of several multi-wavelength campaigns of the M dwarf flare star EV~Lac (Osten et al. 2005, Osten et al. in prep.); grey 
red dotted histogram is the result of several monitoring observations of the ultracool dwarf LP349-25 (Ngoc et al. 2007, Osten et al. 2009); gray dotted 
histogram is the results of a deep centimeter-wavelength catalog of the Orion Nebula Cluster (Forbrich et al. 2016).  
Converting these to probability
distributions enables a more realistic assessment of the likelihood of observing a given radio luminosity, and hence
detectability of these types of radio sources, than averages or maximum luminosities.
}
\end{figure}

\begin{figure}
\vspace*{-1cm}
\hspace*{-1cm}
\includegraphics[scale=0.45,trim=0 0 30 100,clip]{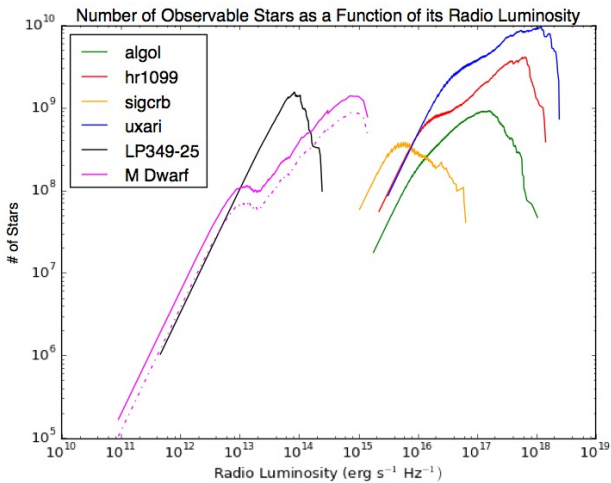}
\vspace*{-3cm}
\caption{ Number of stars detectable with the ngVLA as a function of radio luminosity, using the radio luminosity distributions shown in Fig. 4.
A five minute integration, at a frequency of 10 GHz (check) is assumed.
}
\end{figure}

\begin{figure}
\vspace*{-2cm}
\hspace*{-1cm}
\includegraphics[scale=0.45,trim=0 0 30 100,clip]{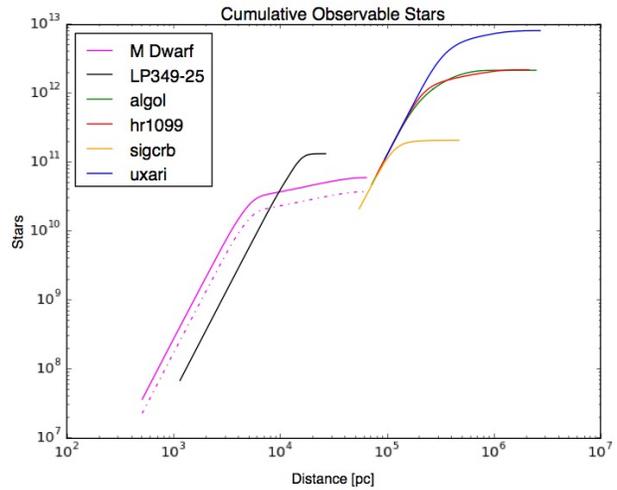}
\vspace*{-3cm}
\caption{ Cumulative number of stars observable as a function of distance. 
}
\end{figure}
\begin{figure}
\vspace*{-2cm}
\hspace*{-1cm}
\includegraphics[scale=0.5]{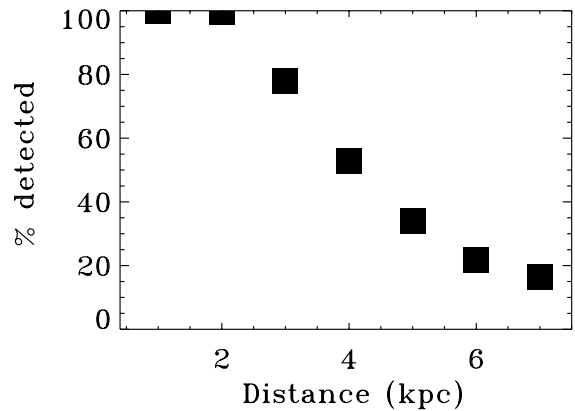}
\vspace*{-7cm}
\caption{Percentage of the 478 radio sources detected in the deep exposure of the Orion Nebula Cluster reported by Forbrich et al. (2016), for clusters
at various distances.  Sensitivities appropriate to a one hour 8 GHz observation with the ngVLA are used.
}
\end{figure}

Our approach for YSOs differed slightly. Since YSOs are found in clusters at discrete distances, 
for this population, we estimated the fraction of sources
which would be detectable in a cluster similar to the ONC but placed further away.  For these purposes we used the ngVLA's continuum rms in one hour
at 8 GHz to set the percent of sources detected for a 3$\sigma$ detection. Figure~7 displays the result of this calculation, and demonstrates that
100\% of the sources could be recovered in clusters at 2 kpc distance, sufficient to characterize many of the typical radio-emitting sources
in star forming regions located in the nearest parts of the Sagittarius and Perseus arms of our galaxy.

\subsection{Constraining Particle Distributions and Magnetic Fields in Magnetic Reconnection Flares } 
Radio observations provide a unique way to characterize the nature
of the accelerated electron population near the star. This is
important for understanding the radiation and particle environment in which close-in exoplanets
are situated.
An aspect of star-planet interactions which can be probed with radio observations 
is the magnetospheric interactions of a magnetized
close-in planet with its host star. Similar to the idea that the proximity of two stellar 
magnetospheres due to close passage can cause periodic episodes of magnetic reconnection (as in Massi et al. 2006),
recent results (Pilliterri et al. 2014) have suggested a triggering mechanism for regularly recurring stellar flares on 
stars hosting close-in exoplanets (hot Jupiters).
For radio wavelength observations,  the time-dependent response of radio emission 
reveals the changing nature of the magnetic field strength and number and distribution of accelerated particles.
As stellar radio emission typically has a peak frequency near 10 GHz, a wide bandwidth system spanning this range
can probe the optically thick and thin conditions, diagnose the changing conditions during the course of a 
magnetic reconnection flare associated with close passage of an exoplanet, and deduce the nature of the accelerated
particle population through measurements of spectral indices from confirmed optically thin emission. This would
provide constrains on the accelerated particle population of close-in exoplanets unavailable from any other
observational method; such a constraint is necessary to perform detailed modeling of the atmospheres of such exoplanets,
due to the influence of accelerated particles in affecting the chemical reactions in terrestrial planet atmospheres 
(Jackman et al. 1990). 

\begin{figure}
\includegraphics[scale=0.35]{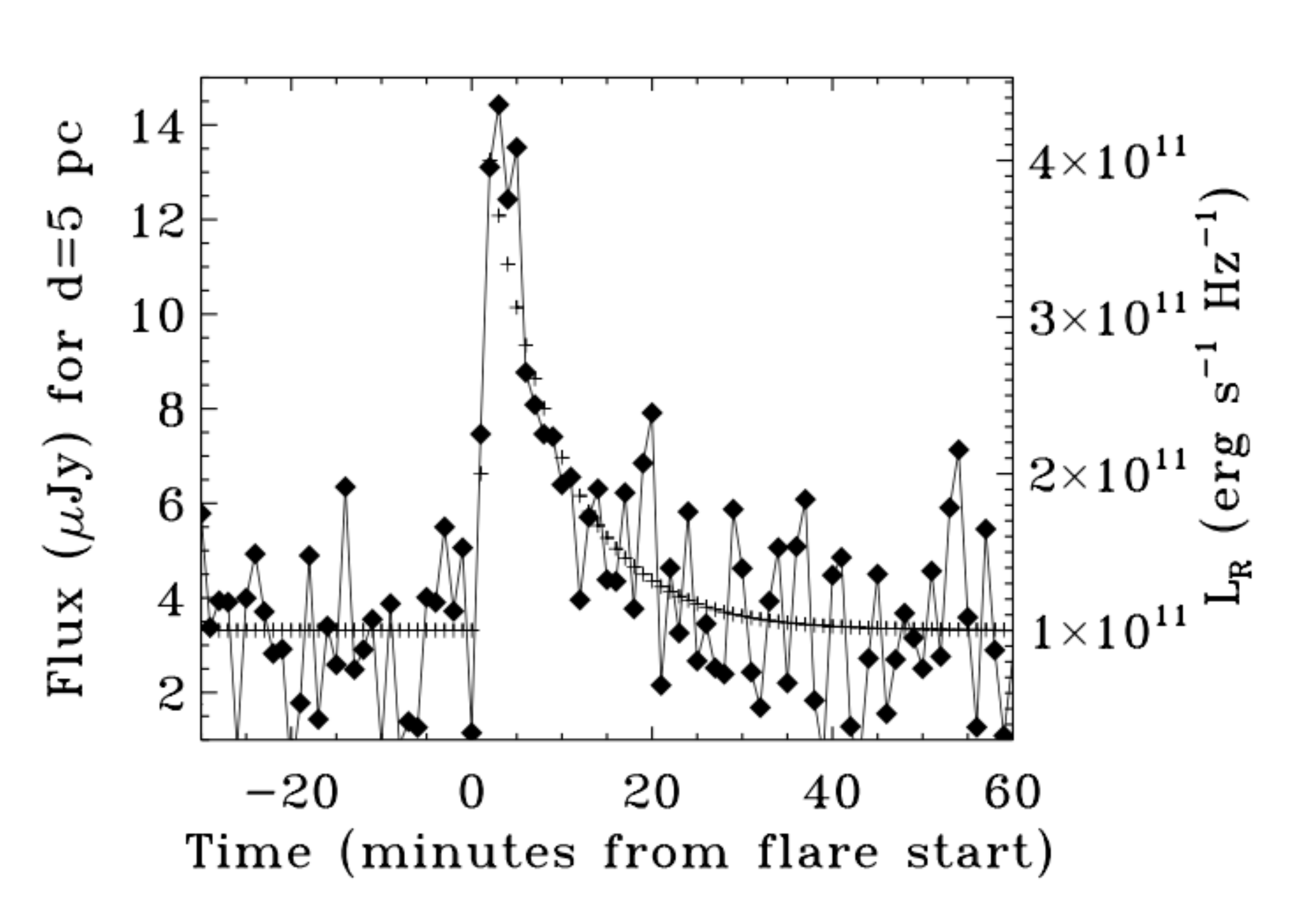}
\hspace*{-1.5cm}
\includegraphics[scale=0.4]{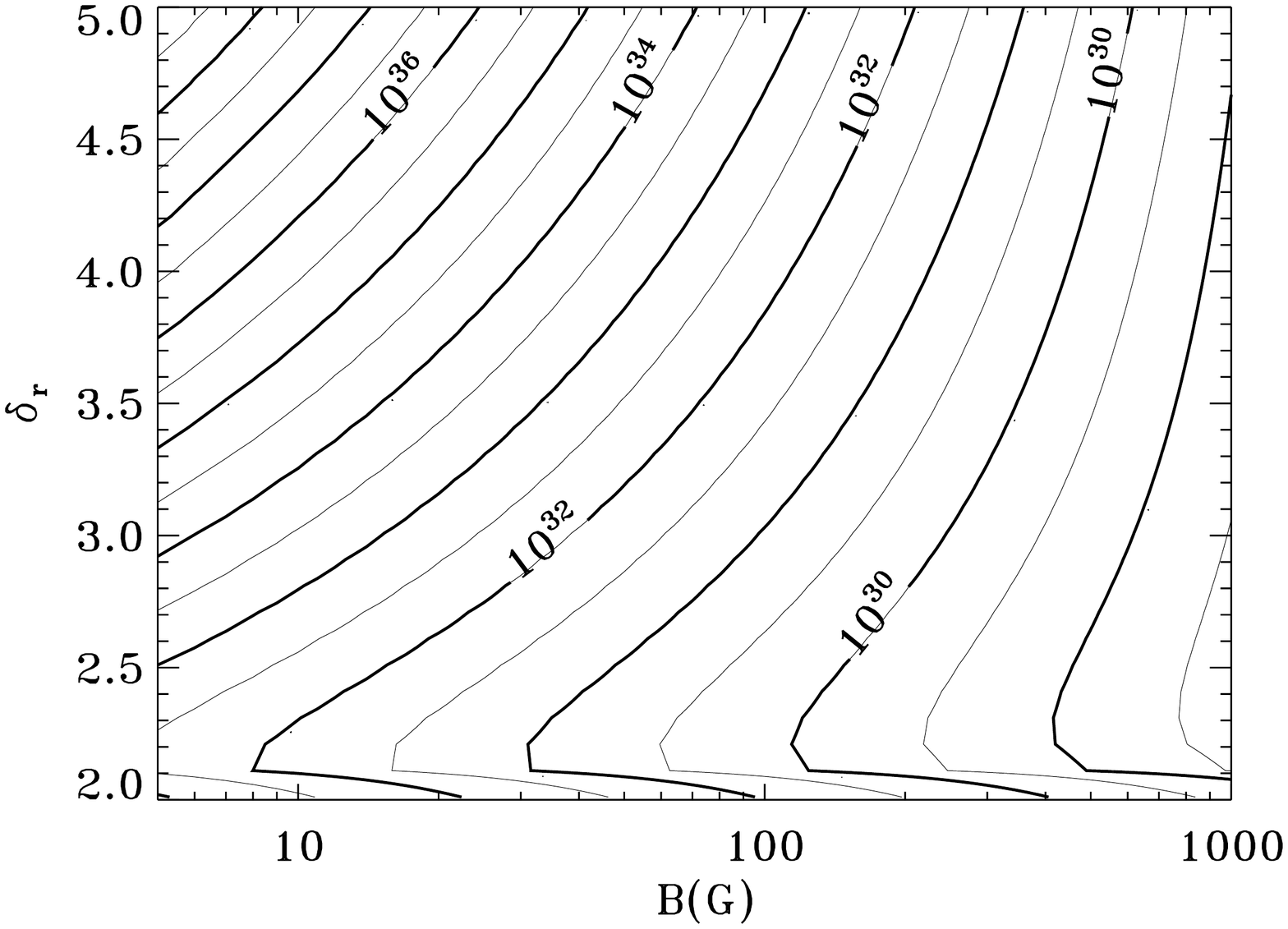}
\vspace*{-1.5cm}
\caption{
(top) Light curve of a short duration (lasting 5 minutes), small enhancement  (factor of three increase
above quiescence) flare on a nearby (5 pc) star. Right yaxis gives intrinsic radio luminosity, left axis
gives the estimated radio flux density at 17 GHz. Noise is estimated in each 1 minute bin by offsetting the
flux by a random number times the expected rms values.
(bottom)
Contour plot of $\delta_{r}$, the index of accelerated particles, and $B$, the magnetic field strength in the
radio-emitting source, for the radio flare seen above.
\label{fig:ptcls}}
\end{figure}

Smith et al. (2005) and Osten et al. (2016) laid out the basic framework for such a measurement, which requires 
multi-wavelength observations of the radio flare and its counterpart at higher energies. 
The uniqueness of the measurement
 lies in the difficulty in determining the characteristics of accelerated particles in nondegenerate environments
 in the cool half of the HR diagram. The usual technique for solar physicists is to deconvolve hard X-ray spectra exhibiting
 optically thin measurements of nonthermal 
 bremsstrahlung emission to determine the power-law index of the accelerated electrons
 and source size. Because of astrophysical detector sensitivity coupled with typically hot thermal flaring plasma, this
 observational method has not yielded results in stellar cases. Radio constraints on accelerated particle characteristics are therefore the only 
 measurement technique that will yield results for planet-hosting stars in the solar neighborhood.

The integrated radio flare energy provides a constraint on the accelerated particle
kinetic energy, under the assumption of optically thin emission. 
The radio light curve provides constraints on  both
the index of the distribution of accelerated particles, as well as the magnetic field strength in the radio-emitting source. 
The contour plot of $\delta_{r}$, the index of accelerated particles, versus the magnetic field strength in the radio-emitting
source as a function of kinetic energy in the accelerated particles in a stellar superflare reported in Osten et al. (2016)
was for an event with a peak radio luminosity at 15 GHz of 5$\times$10$^{16}$ erg s$^{-1}$ Hz$^{-1}$.
 Based on Figure~5, there are numerous M dwarfs detectable with the ngVLA at the lower radio luminosity end, near
 10$^{11-12}$ erg s$^{-1}$ Hz$^{-1}$ for which similar analyses could be performed.  Using the same methodology, we worked out the sensitivity to a radio flare
 with a peak flare enhancement of only 3 times this base level, for a flare lasting only 5 minutes total duration.
 Such an example is shown in the top part of Figure~\ref{fig:ptcls}.
 These types of observations would be most sensitive to the nearest M dwarfs, in order to study time-variable emission
 over very short timescales (minutes). Quiescent radio luminosities as low as 10$^{11}$ erg s$^{-1}$ Hz$^{-1}$ (at the limit of current detection capabilities; Osten 
 et al. 2015) would be detectable in short integrations, allowing for study of small enhancement events only factors of a few larger.
 The kinetic energies probed here are moderate to large sized solar flares, and probe magnetic field strengths
 in the tens to hundreds of Gauss level. 
 
 A wide bandwidth system spanning the 10-20 GHz range can probe the optically thick and thin conditions, diagnose the
 changing conditions during the course of a magnetic reconnection flare associated with the close passage of an exoplanet,
 and deduce the nature of the accelerated particle population through measurements of spectral indices from confirmed
 optically thin emission. Operation of multiple subarrays would enable the frequency coverage available simultaneously to be
 significantly expanded. This would provide constraints on the accelerated particle population of close-in exoplanets
 unavailable from any other observational method; such a constraint is necessary to perform detailed modeling of the
 atmospheres of such exoplanets, due to the influence of accelerated particles in affecting the chemical reactions
 in terrestrial planet atmospheres.
 
 \section{Summary}
 
 The influence that stars have on their near environments is a timely research topic now, and is expected
 only to grow in the future as exoplanet discoveries grow.  While we expect that new discoveries 
 in this area in the next decade will fill in some areas where we currently lack insight, there will remain gaps in our knowledge which can only be filled by a next generation sensitive radio telescope operating in the deka-GHz range.

\section{References}
 \noindent  Anglada-Escud\'{e}, G. et al. 2016 Nature, 536, 437 \\
  Bower, G. et al. 2015 arXiv 1510.06432  \\
  Condon, J. et al. 1998 AJ 115, 1693 \\
  Crosley, M. K. et al. 2017 ApJ 845, 67\\
  Drake, S. A. et al. 1987 AJ 94, 1280 \\
  Drake, J. et al. 2013 ApJ 764, 170\\
  Dressing, C. \& Charbonneau, D. 2013 ApJ 767, 95 \\
   Favata, F. et al. 1995 A\&A 298, 482 \\
    Fichtinger, B. et al. 2017 A\&A 599, 127\\
  Forbrich, J. et al. 2016 ApJ 822, 93 \\
  Garraffo, C. et al. 2016 ApJ 833, L4\\
  Garraffo, C. et al. 2017 ApJ 843, L33\\
Heyner, D. et al. 2012 ApJ 750, 133 \\
  Jackman, C. H. et al. 1990 JGR 95, 7417 \\ 
  Johnstone, C. P. et al. 2015 A\&A 577, 28\\
  Lovelace, R. et al. 2008 MNRAS 389, 1233 \\
  Massi, M. et al. 2006 A\&A 453, 959 \\
  Ngoc, P. et al. 2007 ApJ 658, 553 \\
  Osten, R. A. \& Bastian, T. 2006 ApJ 637, 1016 \\
  Osten, R. A. \& Bastian, T. 2008 ApJ 674, 1078\\
  Osten, R. A. et al. 2005 ApJ 621, 398 \\
  Osten, R. A. et al. 2009 ApJ 700, 1750 \\
  Osten, R. A. et al. 2016 ApJ 832, 174\\
  Osten, R. A. \& Wolk, S. J. 2015 ApJ 809, 79\\
  Osten, R. A. \& Wolk, S. J. 2017 IAUS 328, 243\\
  Panagia, N. \& Felli, M. 1975 A\&A 39, 1\\
  Pilliterri, I. et al. 2014, ApJ 785, 145 \\
  Reid, I. N. et al. 2007AJ 133, 2825 \\
  Reid, I. N. et al. 2008 AJ 136, 1290\\
  Richards, M. et al. 2003 ApJS 147, 337 \\
  Segura, A. et al. 2010 AsBio 10, 751\\
  Smith, K. et al. 2005 A\&A 436, 241\\
  Tian, F. et al. 2014 E\&PSL 385, 22\\
  Venot, O. et al. 2016 ApJ 830, 77\\
  Vidotto, A. et al. 2015 MNRAS 449, 4117 \\
  Wargelin, B. J. \& Drake, J. 2001 ApJ 546, L57\\
  Wood, B. E. et al. 2004 AdSpR 34, 66\\
  Wood, B. E. et al. 2014 ApJ 781, L33\\
  Wright, A. E. \& Barlow, M. J. 1975 MNRAS 170, 41\\

\end{document}